\documentclass[onecolumn,superscriptaddress,amsmath,amssymb,floatfix]{revtex4}

\usepackage{graphicx}% Include figure files
\usepackage{dcolumn}% Align table columns on decimal point
\usepackage{bm}% bold math
\usepackage{textcomp}
\usepackage{color}
\usepackage{setspace}
\usepackage{txfonts}
\begin{document}

\title{Evolving fracture patterns: columnar joints, mud cracks, and polygonal terrain}
\author{Lucas Goehring*}
%\email[]{lucas.goehring@ds.mpg.de}
\address{Max Planck Institute for Dynamics and Self-Organization, Am Fassberg 17, D-37077 G\"ottingen, Germany \\ *lucas.goehring@ds.mpg.de}
%\date{\today}
\begin{abstract}

When cracks form in a thin contracting layer, they sequentially break the layer into smaller and smaller pieces.  A rectilinear crack pattern encodes information about the order of crack formation, as later cracks tend to intersect with earlier cracks at right angles. In a hexagonal pattern, in contrast, the angles between all cracks at a vertex are near 120$^\circ$. Hexagonal crack patterns are typically seen when a crack network opens and heals repeatedly, in a thin layer, or advances by many intermittent steps into a thick layer.  Here it is shown how both types of pattern can arise from identical forces, and how a rectilinear crack pattern can evolve towards a hexagonal one.  Such an evolution is expected when cracks undergo many opening cycles, where the cracks in any cycle are guided by the positions of cracks in the previous cycle, but when they can slightly vary their position, and order of opening. The general features of this evolution are outlined, and compared to a review of the specific patterns of contraction cracks in dried mud, polygonal terrain, columnar joints, and eroding gypsum-sand cements.

\end{abstract}
%\classification{CLASSIFICATIONS} 

\maketitle

\section{Introduction}

\begin{figure}
\includegraphics[width=13.5cm]{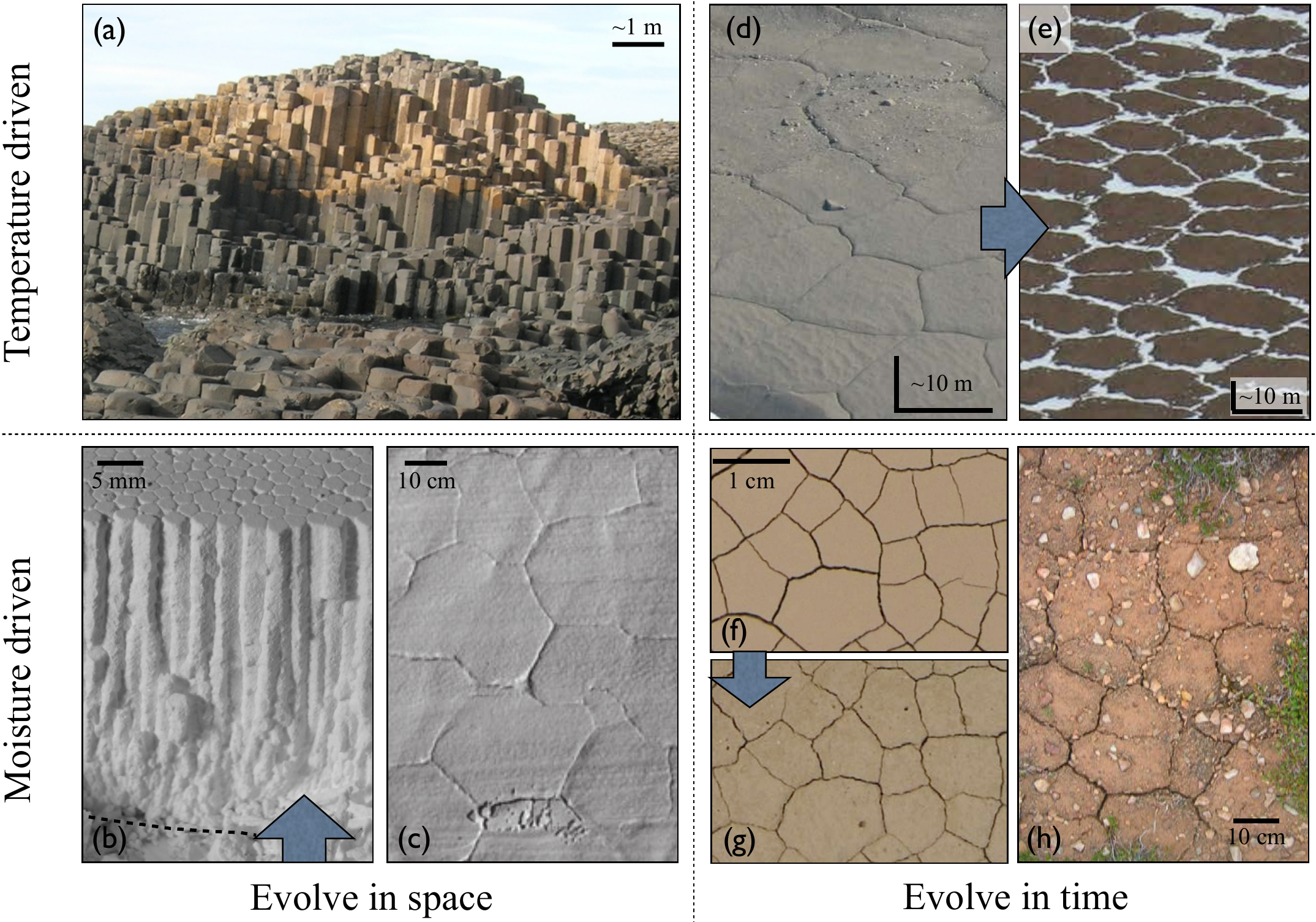}
\caption{\label{cracks} Hexagonal crack patterns are found naturally in many settings. They can be driven by thermal or desiccation contraction, and appear to require the opportunity to evolve, either in space or time.  Shown here are examples of columnar joints in (a) lava at the Giant's Causeway (image courtesy of S. Morris) and (b) in desiccated starch, (c) desiccation cracks in the cemented sands on the windward slopes of dunes (reproduced from Ref. \cite{Chavdarian2010}), (d) `young' thermal contraction cracks, or polygonal terrain, in Taylor Valley, Antarctica, which is speculated \cite{Sletten2003} to be in the process of evolving into a mature pattern such as (e) that in Beacon Valley, Antarctica, and desiccation-induced mud-crack patterns in the lab after (f) one drying and (g) twenty-five cycles of wetting, drying, and cracking \cite{Goehring2010b}, and (h) as found naturally in soil. Images (a,d,e) were taken obliquely, and the scale estimated by feature sizes. In (b) the starch columns are shown inverted, with the original drying surface indicated by a dashed line.}
\end{figure}

Cracks in cooling or drying media can form captivating patterns of connected networks, such as the artistic craquelure patterns sometimes seen in pottery glazes, to those found in dried mud,  or the polygonal networks covering the polar regions of Earth and Mars.  Two types of pattern are common.  The first is a rectilinear pattern, such as may form when an homogeneous slurry is dried or cooled uniformly \cite{Shorlin2000,Bohn2005,Goehring2010b}. The vertices of these patterns typically contain one crack path intersecting another straight crack at right angles, a T-junction.  Secondly are crack networks with a more regular hexagonal pattern, such as the columnar joints of the Giant's Causeway \cite{RBSRS1693,Huxley1881}.   These are dominated by cracks intersecting at 120$^\circ$, or Y-junctions.  Curiously, examples of both extreme types of patterns can sometimes be found in the same systems, such as garden-variety mud-cracks.  These, and other examples, are shown in Fig. \ref{cracks}.  Hexagonally ordered patterns may form as the result of either desiccation or thermal contraction cracks, and there is a deep physical symmetry between these two mechanisms \cite{Biot1941,Norris1992}.  Furthermore, they may be classified into crack patterns that have evolved in both time and space (such as a network of crack tips that delimit the prismatic forms of columnar joints as they advance \cite{Mallet1875}), and those that have recurred in the same place, but evolved over time as the pattern repeatedly healed, and re-cracked \cite{Lachenbruch1962,Goehring2010b}.  

Here it is shown that both hexagonal and rectilinear crack patterns can arise naturally from the same assumptions of fracture behaviour, and how a rectilinear, T-junction dominated pattern can develop into a hexagonal pattern, with Y-junctions.  This paper is roughly divided into two parts.  In the first part (Sections 2,3), the relevant physics of linear elastic fracture mechanics in a thin brittle layer is summarised, and used to develop a model for the evolution of a rectilinear crack pattern into a hexagonal one, through the cyclic opening of cracks.  The second part (Sections 4-7) reviews the ordering of crack patterns in a number of geophysical systems: desiccation cracks in clays that are repeatedly wetted and dried; the thermal contraction cracks in permafrost, known as polygonal terrain; columnar joints in lava and starch; and cracks in eroding gypsum sand.  These patterns are each explored, and shown to obey the same general physical mechanism of crack pattern evolution.  In the case of desiccation cracks in clays new experiments are also reported, which show that the length-scale of  the evolution of a rectilinear mud-crack pattern towards a hexagonal one scales linearly with the thickness of the cracking layer.  

\section{Contraction cracks in thin layers}

\begin{figure}
\includegraphics[width=13.5cm]{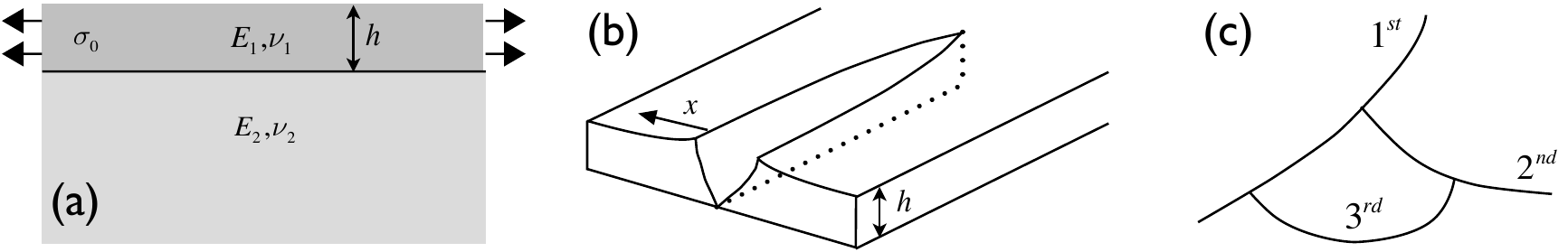}
\caption{\label{model} Schematic of thin film fracture.  (a) A film of height $h$ under stress $\sigma_0$ and with elastic properties $E_1$ and $\nu_1$ is adhered to a much thicker substrate with potentially different elastic properties $E_2$ and $\nu_2$. (b) In such a film a channel crack may open and grow across the film.  The crack releases stress preferentially in the direction perpendicular to its growth.  (c) If many such cracks form, they will interact with each other.  Later cracks will curve to intersect earlier ones at right angles, the direction that maximises the strain energy release rate as they grow.}
\end{figure}

When an elastic body cools or dries it tends to shrink, and may crack.  Here the case of a brittle film under tensile stress, and firmly attached to a much thicker, non-cracking substrate, as sketched in Fig. \ref{model}(a), is reviewed.  As a drying body looses water, it develops an internal pore pressure as a result of capillary effects.  In the absence of cracks the equations of stress equilibrium and poro-elasticity are easily solved, and predict an in-plane stress, $\sigma_0$, proportional to this capillary pressure.  Similarly, if a film is cooled, it will have an in-plane stress proportional to the temperature change, and to the mismatch in thermal expansivities of the film and substrate.  The similarity between the patterns of cracks that develop during drying and cooling reflects the underlying similarity between the physics of poro-elasticity, and thermo-elasticity \cite{Biot1941,Norris1992}.  

Tension is released when a crack opens in the brittle layer, as the normal stresses in the film must vanish on the new crack surfaces. The substrate will resist deformation, however, and exert a restoring force on the film-substrate interface, approximately proportional to the local in-plane strain \cite{xia2000,Yin2008}.  For a long straight crack in a thin film, as shown in Fig. \ref{model}(b), these conditions give rise to an exponential dependence of the height-averaged film stress on the distance $x$ away from the crack,
\begin{equation}
\label{stress}
\sigma_\perp = \sigma_0(1-e^{-x/l}), \>\>\> \sigma_\parallel = \sigma_0(1-\nu_1e^{-x/l}) 
\end{equation}
where $\sigma_\perp$ and $\sigma_\parallel$ are the stress normal and parallel to the crack face, and $\nu_1$ is Poisson's ratio of the film \cite{xia2000}.  The length-scale over which stresses relax, $l$, is proportional to the film thickness, $h$, 
\begin{equation}
\label{stresslength}
l = \frac{\pi}{2}hg(\alpha,\beta).
\end{equation}
The order-one term $g$ is given numerically in reference \cite{Beuth1992} and depends on the Dundurs parameters $\alpha$ and $\beta$, ratios involving the plane-strain elastic moduli $E_1$ and $E_2$ and Poisson ratios $\nu_1$ and $\nu_2$ of the film and substrate.  If there is no elastic mismatch between the cracking layer and its substrate, $l$ = 2.0 $h$, while for a compliant film on an un-yielding substrate $l$ is between 1.2 and 1.3 $h$, depending on $\nu_1$.   In the limit of very soft (or fluid) substrates, $l$ diverges, and the film behaves as an unconstrained layer. 

Once a crack has formed, the Griffith criterion compares the energy required to create new fracture surface, $G_c$, to the rate of change of elastic energy in the cracked body for an infinitesimal extension of the crack, $G$, and predicts crack growth when more energy would be released, then consumed \cite{Lawn1993}.   While the first term, the critical energy release rate, is a material parameter, the second term depends on the geometry of the crack and its local surroundings.  For an isolated straight film crack that is long compared to the film thickness \cite{Beuth1992,xia2000}, the strain energy release rate
\begin{equation}
\label{G}
G = \frac{\pi}{2} \frac{h \sigma_0^2}{E_1}g(\alpha,\beta) = \frac{\sigma_0^2l}{E_1}.
\end{equation}
Thus, when the film stress reaches some critical value $\sigma_c = \sqrt{{EG_c}/{l}}$, the film can crack.  
Since $G_c$ is a material constant, thinner layers will require higher stresses to crack than thicker layers, a fact well-established in the engineering literature \cite{Hutchinson1992,Chiu1993a,Thouless1995}.

As stress develops, it can be relieved by the sequential fragmentation of the film. This may be illustrated by considering an array of parallel cracks, each separated by a distance $\lambda$ that is large compared to $l$.  If each crack relaxes stress as per Eqn. \ref{stress}, then the energy release rate, per crack, of the array is 
\begin{equation}
G =G_c\frac{\sigma_0^2}{\sigma_c^2}\tanh\bigg(\frac{\lambda}{2l}\bigg),
\end{equation}
where $\sigma_c$ is the stress required to open a single isolated crack, or critical cracking stress \cite{xia2000}.  As drying or cooling proceeds, the background stress $\sigma_0$ increases, and the allowed crack spacing $\lambda$ will shrink. Counter-intuitively, this process does not continue indefinitely, no matter how high the film stress gets. When the cracks are closer than a few film thicknesses to each other the stresses are no longer well-approximated by a through-thickness-average stress.  Numerical simulations of the stress state between two parallel cracks have shown that, for cracks closer than some limiting spacing $\lambda_c\sim h$, the surface stress in the region between the two cracks is compressive, and will tend to close any further flaws or potential cracks \cite{Bai2000,Yin2010}.  For cracks in a compliant layer on a hard substrate, experiments have confirmed that the crack spacing is limited to about 3 times the film thickness \cite{Allain1995,Hull1999,Shorlin2000,Pasricha2009,Goehring2010b}.  It is likely that in other cases the saturation spacing scales with the stress relaxation length $l$, although this has not been rigorously demonstrated.  If so, then where there is no difference in the elastic properties of the cracking layer, and its substrate, the limiting crack spacing should be  $\sim$5 times the depth of the cracks.  

When multiple cracks form they will interact with each other and pattern the film. Rather than the simple parallel cracks just considered, cracks will normally nucleate randomly on material defects or along the faces of pre-existing cracks and grow until they hit something, such as a boundary or other crack.  Each crack will tend to grow in the direction that maximises $G-G_c$, the difference between the strain energy released, and the fracture energy consumed, during growth \cite{Lawn1993}.  This is a local principle, as the energy release rates are calculated {\it at the point and time} of the crack growth.  As in Eqn. \ref{stress}, any pre-existing crack is more efficient at relieving stress perpendicular to its face, than parallel to it.  Therefore, a growing crack will tend to curve to intersect another crack at right angles to it.  As shown in Fig. \ref{model}(c), the resulting T-junctions thus encode time information that can be used to partially reconstruct the order in which cracks form \cite{Bohn2005}.

\section{A model for evolving crack patterns}

Cracks in a homogeneous film will tend to pattern the surface into roughly equally-sized rectilinear pieces, bounded with right-angled T-junctions.  However, many natural crack patterns, such as columnar joints, show hexagonal planforms, and equi-angled Y-junctions.  In both cases Euler's formula implies that the average number of neighbours for any polygon is 6, as long as cracks do not cross each other, or vertices form from more than three cracks.  It is possible to continuously deform a rectilinear crack network into a hexagonal one (allowing vertices to pass each other, if necessary \cite{Goehring2005}), if the pattern is allowed to evolve.  Development towards a hexagonal pattern might thus be expected by noting that a hexagonal tiling minimises ratio of the crack-length (perimeter) to area of the polygons \cite{Mallet1875}.  However, as discussed in the previous section, a growing crack tip can only respond to the local strain energy release rate at position and moment where it is opening.  There is no general way to link this local energy maximisation principle to a global constraint.  Here, instead, it is shown that evolution from a rectilinear to hexagonal pattern can be explained by entirely local responses of advancing cracks, as the result of meeting three conditions: (1) that cracks recur or advance repeatedly, (2) that the previous positions of cracks act as lines of weakness, guiding the next iteration of cracking; and (3) that the order of opening of cracks may change in each iteration.  

In order to develop from a rectilinear network of cracks into a hexagonal one, the pattern must have some means of changing with time.  Polygonal terrain consists of networks of thermal contraction cracks that open during the cold polar winters, and which thaw, and heal, during summers; the crack pattern reforms each year \cite{Lachenbruch1962}. For columnar joints the pattern of growing cracks at any particular moment is confined to a thin region, approximating a 2D film \cite{Goehring2009}.  Each crack advances step-wise, and the columns are the forms delimited by the planar crack network as it advances through space.  

Even if a layer cracks and heals repeatedly, the pattern will not necessarially evolve.  If the system has no memory of previous patterns, each cycle will lead to a novel, but statistically similar, rectilinear crack pattern.  If there is memory of vertex positions only, perhaps through fixed defects ({\it e.g.} a pebble in a mud puddle), then cracks may still nucleate along the same points in each generation, but then grow to intersect with each other at right angles, again creating a statistically similar pattern each time they open.  If any memory is too strong, however, the crack pattern may simply repeat in each cycle without change.  To evolve away from a rectilinear pattern new cracks must be guided, but imperfectly, by previous cracks.

A nascent crack can be guided by a crack in an earlier cycle in a number of ways.  During columnar joint formation each crack advance follows the crack above it, which concentrates stress along its tip, but can grow downwards in a slightly different direction \cite{Aydin1988}.  In a thin layer, the previous cycle's crack may not have fully healed, or may have damaged their vicinity.  In either case the critical energy release rate $G_c$ is lower near old crack paths \cite{Goehring2010b}.  Alternatively, if the previously cracked region is more compliant (lower $E_1$) than uncracked material, then near a previous crack the strain energy release rate $G$ will be higher, as suggested by the form of Eqn. \ref{G}.  Finally, as with polygonal terrain, the crack may leave a depression after healing \cite{Lachenbruch1962,Sletten2003}, locally lowering the average height $h$ by $\delta$.  In this case the total energy spent by a unit length of crack extension would  be reduced to $G_c(h-\delta)$.  However, the rate of energy release, $G$, remains integrated throughout the region surrounding the crack of width $\sim h$, and will be relatively unchanged. It will thus be energetically favourable for a crack to be confined within the depression of the former crack.  The action of any, or a combination, of these factors means that a crack tip advancing to maximise $G-G_c$ will be guided to follow a previously existing crack path as it grows. 

\begin{figure}
\includegraphics[width=13.5cm]{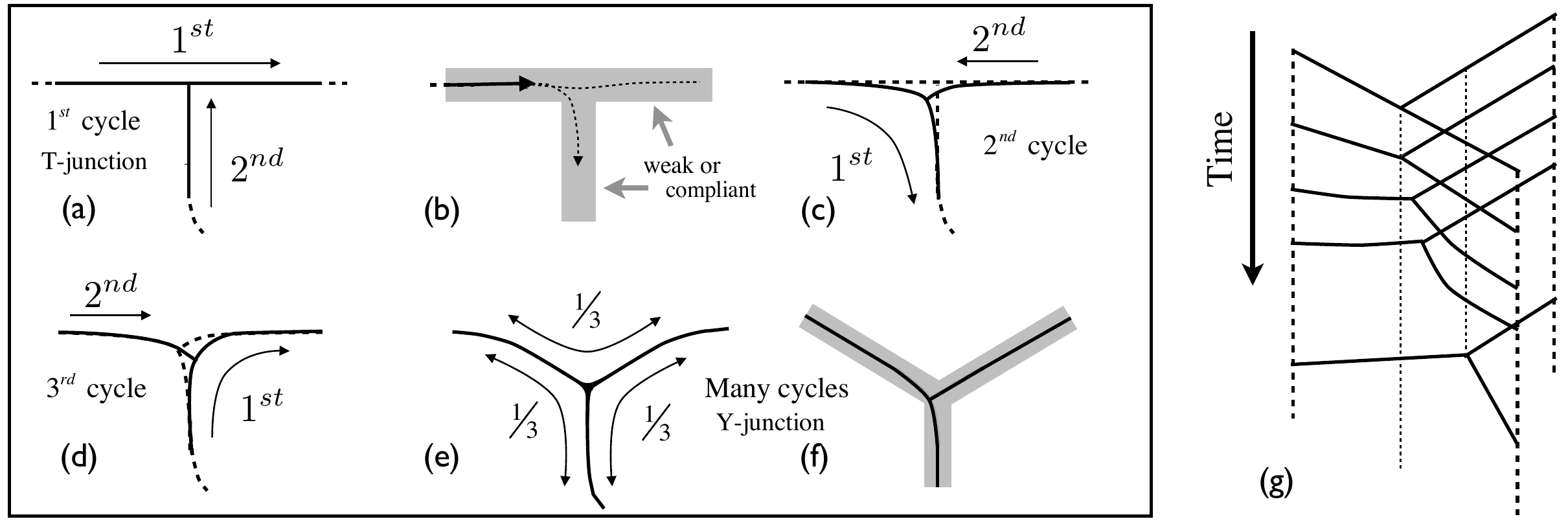}
\caption{\label{evolve} Schematic model for the evolution of a rectilinear crack pattern into a hexagonal pattern.  (a) Initially, the order of crack opening is captured by the shape of the T-junctions.  (b) If the cracks heal and recur a new crack can be guided by the previous cycle's cracks, but deflected by the asymmetric stress concentration it feels near a vertex, in which case (c) the vertex is shifted, and the crack paths are curved.  (d) This process may repeat, and evolve the crack pattern.  Over many cycles (e), with cracks equally likely to approach a vertex from any branch, symmetry will lead to a Y-junction.  (f) Close to the vertex, however, the order of cracking in any particular cycle will still curve cracks slightly.  (g) If the evolution of the crack network is stacked in space, it provides a model for extended patterns such as columnar joints.}
\end{figure}

Changes in the order at which cracks approach and leave vertices allow for the above conditions to evolve a crack pattern.  The case of a crack guided by a network of partially healed cracks is demonstrated in Fig. \ref{evolve}(a-f).   In this example, consider a film that cracks and heals regularly, but where the regions near previous cracks are more compliant than their surroundings.  In the first generation, as sketched in Fig. \ref{evolve}(a), one crack has advanced from left to right, while a later crack has curved in from below to intersect it at right angles. After healing, in the next cycle of cracking, a crack is guided along the path of a previous crack. As it approaches the area near the previous vertex the stress field around its tip becomes asymmetric ({{\it i.e.} the shear stress intensity factor is non-zero). As has been recently demonstrated, a crack path is attracted by an inclusion of more compliant material \cite{Jajam2012}.  If this attraction is weak, the crack will follow close to its original path, and the pattern will have no opportunity to evolve.  However, as shown in Fig. \ref{evolve}(b), the asymmetry may cause the  crack may be attracted downwards, and curve to follow a different path to its first incarnation.  A further crack approaching the vertex in this cycle, as in Fig. \ref{evolve}(c), will also be deflected downwards slightly, as it curves to intersect the first crack.  Since variations in film stresses can only be felt over distances of a few film thicknesses (Eqn. \ref{stress}), these changes in crack paths will initially only be expected in a small region near the vertex, whose size should vary in proportion to $l$, and thus $h$.  Within this region the new crack paths are deflected, and the angles at which the cracks approach the vertex will be more equi-angled than the original T-junction.

If there are repeated changes in the order at which cracks approach a vertex, then changes in the vertex position will accumulate over many cycles, and lead to a directed drift of the vertex in the direction of the crack that originally approached it at right angles.  For example, if in any cycle a crack approaches first from the lower branch of the vertex, as sketched in Fig. \ref{evolve}(d), the path near the vertex will be unstable, in that any slight perturbation would cause it to deflect to either the left or right branch.  If the right branch is chosen, the crack will further round the vertex, and move it further down.  Over many cycles, if all branches near a vertex crack first with equal frequency, then by symmetry the vertex will stabilise with three equal angles of 120$^\circ$, as sketched in Fig. \ref{evolve}(e).  However, as shown in Fig. \ref{evolve}(f), the order of cracking in any particular cycle will give rise to deviations in the immediate vicinity of vertices, where cracks curve slightly to intersect each other.  Alternatively, this might be noticeable by a damage zone, or through crack tip splittings, near vertices.

A pattern which evolves as it passes through space, such as columnar joints, can be thought of as the structure formed by stacking evolving crack networks over many cycles, as shown in Fig. \ref{evolve}(g).

This model of crack ordering makes several predictions.  First, that any such hexagonal pattern must have cracks recurring repeatedly, near older cracks, but not always opening in the same direction - and, equivalently, that the order at which cracks arrive around any vertex must change with time.  It implies that any environment with a hexagonal crack pattern should also be able to support, and have previously shown, a rectilinear one.   Both types of pattern should scale in the same manner, with crack spacings a few times the stress relaxation length $l\sim h$, as long as sufficient stress exists to saturate the spacing.  The hexagonal patterns must have had a chance to evolve, over many crack opening cycles.  The reverse, that rectilinear patterns are `young', is not necessarily true, as a crack pattern may heal sufficiently between openings to erase any memory, and reset the pattern.  Finally, intermediate patterns should link the rectilinear and hexagonal forms.  The effects of an evolving pattern should be first noticeable by small crack deflections in the vicinities of vertices, which will tend to equalise the angles between cracks.  Further from the vertices, in the early stages of evolution, cracks should still appear to be approaching as like a T-junction.  During this time, the vertices themselves should also show a directed drift in the direction of the cracks that originally approached each vertex at two right angles.

The remaining sections of this article will show how the model conditions and predictions outlined here are realised for desiccation cracks in soils, polygonal terrain, columnar joints, and eroding gypsum sand cements.

\section{Mud cracks}

Desiccation cracking in thin elastic layers, such as clays or other granular media, is a simple model system that is frequently used to study crack patterns. Experiments in thin starch layers have been used to show how the timing information of a crack pattern is encoded in the shape of T-junctions \cite{Bohn2005}, for example, while the vibrating of pastes prior to drying has been used to show how crack patterns are influenced by anisotropy in film properties \cite{Nakahara2005,Nakahara2006}.  In nature, mud cracks can be found with both rectilinear and hexagonal (Fig. 1(h)) cells.   It has been demonstrated experimentally that mud-crack patterns evolve, when they are repeatedly moistened and dried \cite{Goehring2010b}.  Over only a few cycles of cracking and healing, the crack angles approach 120$^\circ$, and vertices move.  In this section, these experiments are further developed, in an attempt to look at how the evolution of mud-crack patterns scales with film thickness. 

Slurries of bentonite clay (Acros Organics, Bentonite K-10) were prepared by mixing one part bentonite with 1.9 parts deionized water (Millipore) by weight, stirring the mixture to be homogeneous, and then immediately pouring into large flat-bottomed glass Petri dishes (diameter 185 or 195 mm, height 15 mm).  The slurries were dried 60 cm under a 500 W halogen heat lamp in a well ventilated room, and were fully dry within 48 hours.  Previous work \cite{Goehring2010b} shows that in this system the crack spacing is about 3 times the layer thickness, except in layers thinner than $\sim$2 mm, when the critical layer thickness is approached.   After each drying phase was complete, the dishes were imaged by a digital camera (Nikon D5100).  Markings made around the circumference of the dishes were used as fixed points to adjust the magnification, position, and rotation of each image to match the configuration of the first drying's image.  Typically images could be matched to within 2 pixels, or 0.1 mm accuracy.  Each dish was then rewet, by spraying with a light mist of deionized water, until it was visibly damp throughout the entire layer (a mass of water between 1.1 and 1.3 times that of the dry clay was added), and then set back to re-dry.

\begin{figure}
\includegraphics[width=13.5cm]{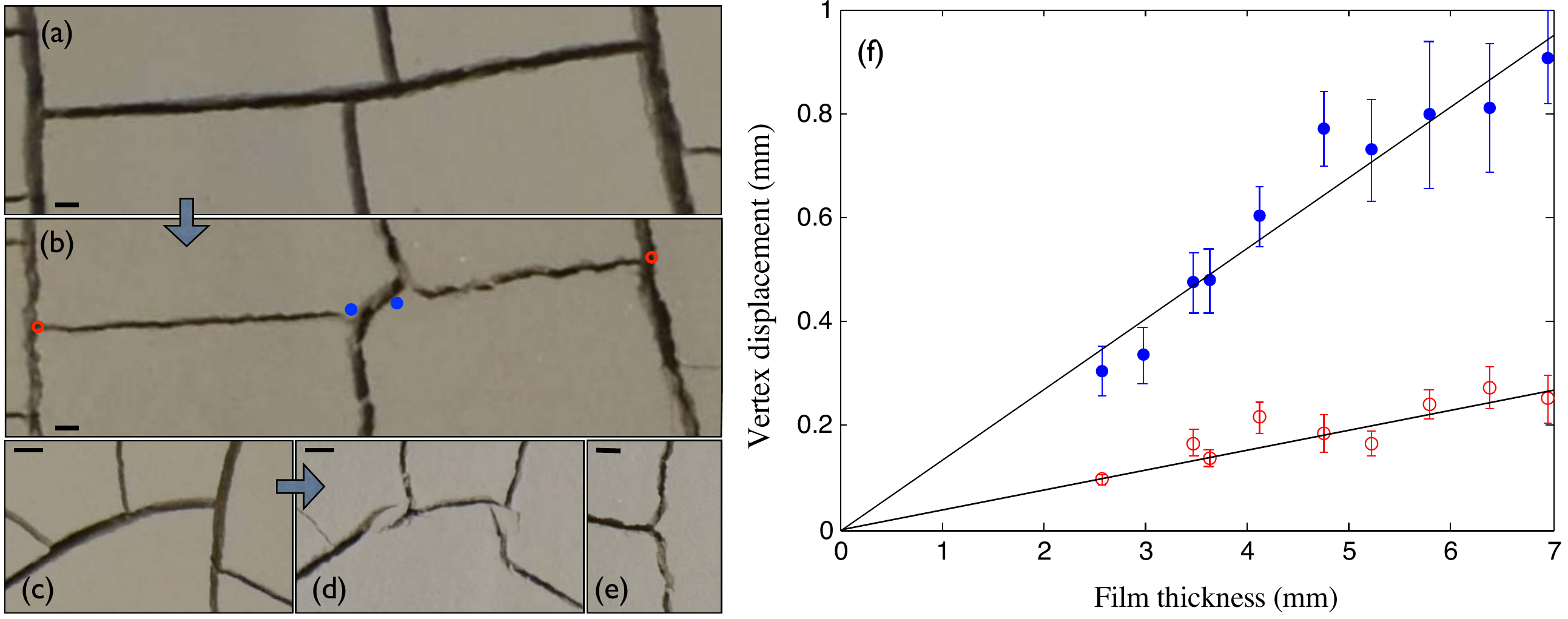}
\caption{\label{mudcracks} Mud-crack patterns change between (a) their first and (b) second dryings. In (b), two vertices have changed the order in which cracks have approached a vertex, and shifted their position noticably as a result (their positions in the first cycle are shown by blue points), while two vertices (red points) have remained relatively undistorted.  In thicker layers, branching of cracks is sometimes noticed, for example between (c) the first and (d) second drying, alongside (e) smoother vertex deflections.  (f) The average displacement of vertices with cracks that have curved away from their original position (blue discs)  and vertices which have not (red circles), are both consistent with a linear dependence on layer thickness.  In (a)-(e) scale bars are 1 mm.}
\end{figure}

\begin{figure}
\includegraphics[width=13.5cm]{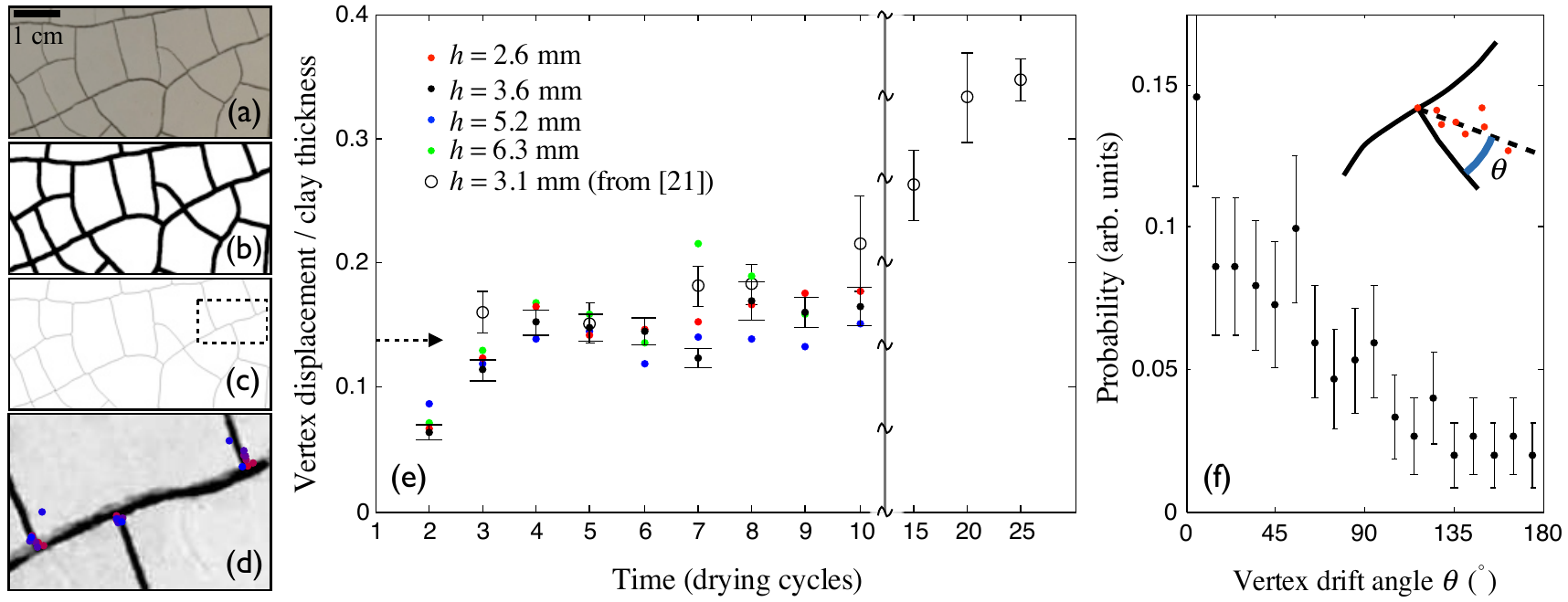}
\caption{\label{motion} Crack vertices move during repeated drying cycles.  To measure this, (a) raw images are (b) thresholded, (c) skeletonized, and then (d) vertex positions are found as the branch-points of the skeletons.  Shown here are the positions of three vertices over ten drying cycles, graded in colour from the first (red) to the tenth (blue) cycle.  (e) The average vertex displacement increases with the number of drying cycles, and scales with the layer thickness.  Representative error bars are given for $h=3.6$ mm.  The data from a previous study \cite{Goehring2010b} where $h = 3.1$ mm are also shown. Vertex motion is most apparent in the first four cycles, and then slows down to a gradual drift. The arrow shows the measured vertex displacement when cracks round a vertex, from Fig. \ref{mudcracks}(f), for comparison.  (f) The vertex motion is directed towards the crack that originally approached a vertex at two right angles.  The angle $\theta$ here is the difference between the direction in which a vertex has been displaced between the first and tenth cracking cycles (for $h$ = 2.6 mm), and the angle of approach of this crack, as demonstrated by the inset.}
\end{figure}

For the first two drying cycles, the positions of vertices were measured by hand from digital images, and compared in 10 dishes between 2.6 and 6.9 mm thick.  Two types of vertices were investigated: those where cracks had deflected around a vertex, and those where cracks had formed as in the first cycle.  Examples of both types are shown in Fig. \ref{mudcracks}(a,b).  The morphology of deflected vertices was sometimes variable.  In the thicker layers, especially, there were cases of branching or {\it en passant} \cite{Fender2010} cracks near vertices (Fig. \ref{mudcracks}(c-d)) alongside more smoothly curving intersections (Fig. \ref{mudcracks}(e)).  For vertices where cracks were deflected off their original path in the second drying cycle, the deflection scaled linearly with the film thickness, as shown in Fig. \ref{mudcracks}(f).   Although these displacements showed large standard deviations, the mean displacement over many such vertices was 14$\pm$1$\%$ of $h$.  For the remaining vertices, there was a smaller average deflection.  However, this was above the limiting resolution of 0.1 mm, and scaled with film thickness, at 5$\pm$1$\%$ of $h$.

\begin{figure}
\includegraphics[width=13.5cm]{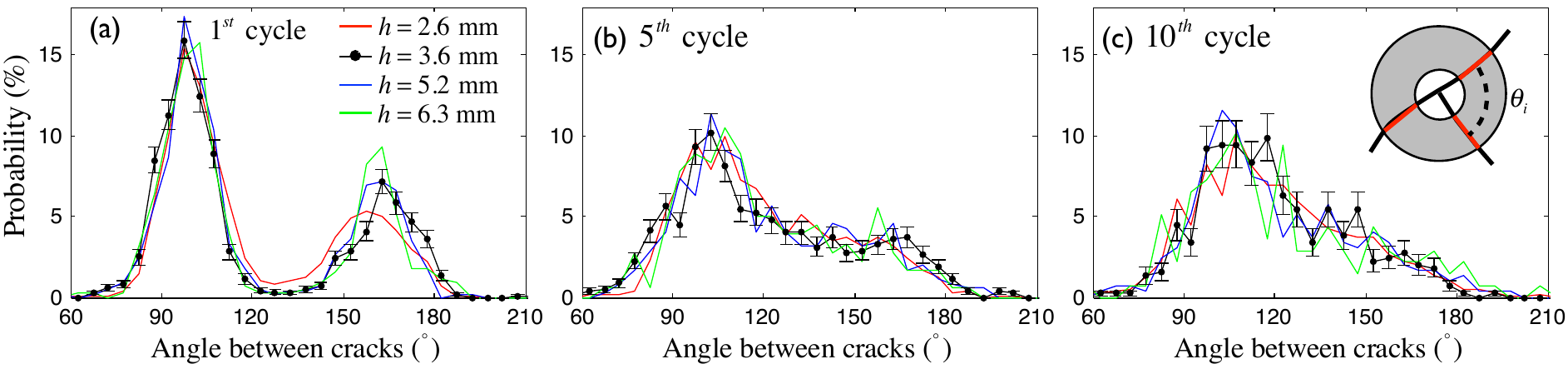}
\caption{\label{angles} The angles at which cracks approach vertices in drying mud-crack patterns move toward the 120$^\circ$ angles of Y-junctions over a few cycles of cracking and healing.  Shown are the probability distribution functions for the angles between cracks of bentonite layers after their (a) first, (b) fifth, and (c) tenth drying is complete.  Representative error bars are shown for the 3.6 mm thick film.  Crack angles are measured between the directions of the lines connecting the crack paths intersecting circles at radii of 5 pixels, and half the cracking layer thickness $h$.  The inner circle is used to minimise the effects of errors in the vertex position due to skeletonization on the angle measurement. }
\end{figure}

For 4 dishes, between 2.6 and 6.3 mm thick, the evolution of the crack pattern was monitored over 10 wetting and drying cycles.  Although it was originally intended to observe for longer, the spray bottle used for rewetting was replaced in the 11$^{th}$ cycle, and the new spray was sufficiently hard to dramatically modify the pattern.  For the first 10 cycles of drying, images of the crack patterns were processed automatically (Fig. \ref{motion}(a-d)) by thresholding greyscale images based on local intensity and gradient (edge detection).  These images were then skeletonized, and vertices found as the branch-points of the thresholded image skeleton.  Using a Matlab-based particle tracking code \cite{Crocker1996}, the motion of the vertices was followed. As with previous experiments \cite{Goehring2010b}, cracks could sometimes disappear for several cycles, only to reappear close to their original location.  For this analysis, only vertices that were present in all 10 crack patterns were analysed.  After this time the average vertex has moved between about 0.5 to 1 mm, depending on the layer thickness.  As shown in Fig. \ref{motion}(e), when scaled with the layer thickness $h$, the motion from all patterns collapse onto a single curve, which also overlaps data published previously \cite{Goehring2010b} for bentonite clay 3.1 mm thick.  In all cases, there was a significant change in the average vertex positions within the first 3 or 4 cycles, followed by a much more gradual vertex motion.  Previously \cite{Goehring2010b}, it was observed that 25-35$\%$ of vertices changed order of formation in any particular generation.  The slowdown in vertex motion thus appears to be identifiable with the time it takes for all vertices to change order of opening once.  Furthermore, the average displacement after 4 cycles, 15$\pm$1\% of $h$, agrees with the vertex shift, shown in Fig. \ref{mudcracks}(f), for the subset of cracks which changed the order of arrival at a vertex between the first and second drying cycles. 

The angle at which cracks approached vertices was also measured, automatically.  An annulus was drawn around each vertex of the skeletonized image, and the points at which the cracks crossed the boundaries of this annulus were found.  An inner radius of 5 pixels was used to minimise the effects of any errors in determining the vertex position precisely.  An outer radius of half the layer thickness was used, to observe the local changes of crack angles near the vertices, which should scale with $h$.   As shown in Fig. \ref{angles}, there is a gradual but systematic equalisation of the crack angles towards 120$^\circ$ over many cracking cycles.  This trend agrees in all experiments.

Finally, Fig. \ref{motion}(f) shows that the direction of the displacement of vertices between the first and tenth drying cycles is well-correlated with the direction of the crack that originally approached the junction at the smallest angles to its neighbours.  In other words, as predicted in Section 3, the vertex is moving preferentially along the direction required to stretch a T-junction into a Y-junction, the direction of the crack that originally approached the vertex at near two right angles. 

\section{Polygonal terrain}

Polygonal terrain consists of networks of thermal contraction cracks, typically a few tens of meters apart, which form in ice-cemented soils \cite{Lachenbruch1962}. It is common in terrestrial circumpolar regions.  The thermal stresses arising from the intense winters of these regions are sufficient to open cracks a few centimeters wide, and several meters deep \cite{Lachenbruch1962,Mackay1974,Mackay2000}. While open, loose soil, snow, and wind-born detritus may fall into the crack. Thus the crack does not heal completely when thawed, and is likely to reopen in the following winter.  The resulting patterns are similar to evolving mud crack patterns, where each seasonal thermal cycle is equivalent to one cycle of drying and wetting. However, the net addition of material into cracks creates wedges of ice or sand that grow at rates of order a millimetre per year and lead to a slow turnover of the near-surface soil \cite{Berg1966,Mackay2002,Sletten2003}. 

Recently, perhaps as a result of the landings of Phoenix \cite{Mellon2009} and Curiosity \cite{Hallet2012} on Martian polygonal terrain, much work has gone into analysing polygonal crack patterns {\it via} satellite imaging.  Large datasets have been collected for sites on Earth and Mars \cite{Ulrich2011,Haltigin2010,Pina2008,Pina2012}. These data have been used to demonstrate that polygonal terrain shows some statistical trends that are also found in foam networks \cite{Pina2008} and that the vertex positions contain spatial correlations \cite{Haltigin2010}.  Work characterising these networks is ongoing and interesting.  Comparisons with ground observations have shown, however, that very high-resolution images ($\sim$0.25 m/pixel) are required to correctly measure the features of polygon vertices \cite{Haltigin2010,Pina2012}, making detailed interpretation of satellite images difficult.  

\begin{figure}
\includegraphics[width=13.5cm]{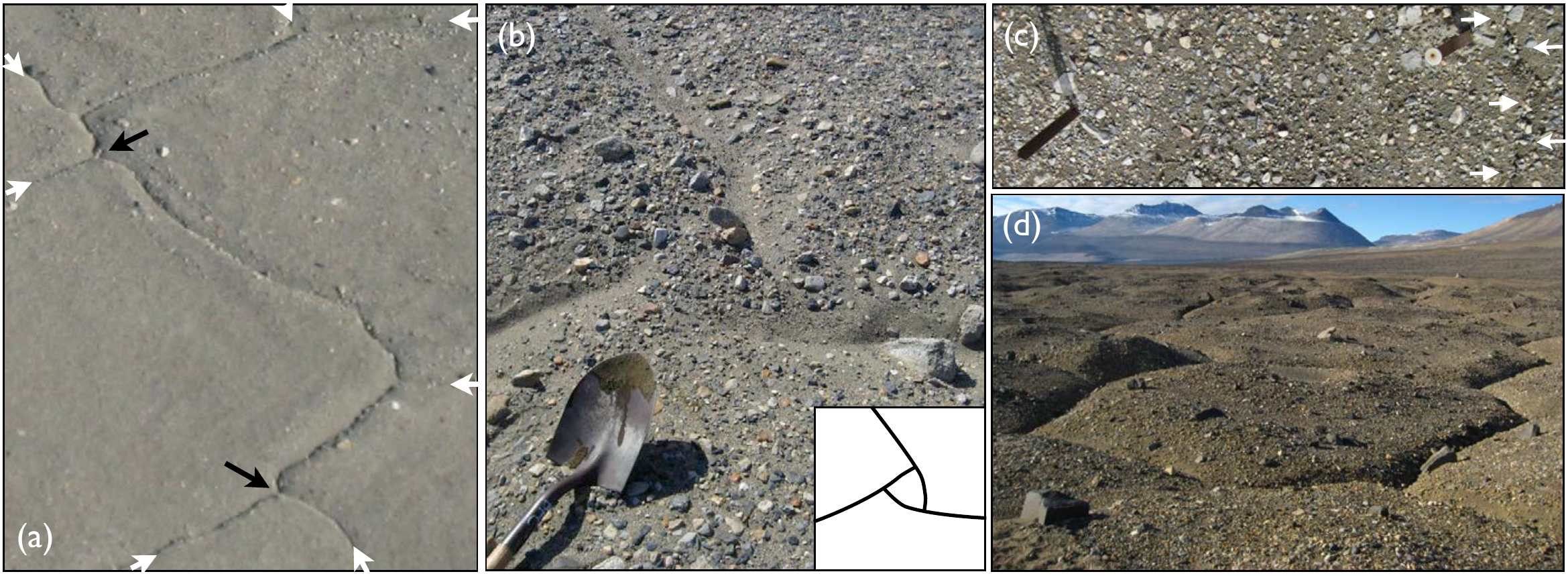}
\caption{\label{polygons} Polygonal terrain in the Dry Valleys, Antarctica.  (a) Oblique/arial view of cracks near New Harbor delta, of age 1 - 2 ka \cite{Sletten2003}.  Crack positions are highlighted by white arrows along the perimeter of the image.  The vertices shown are distorted, and appear to be deflected in the direction indicated by the black arrows from what would otherwise be T-junctions.  (b) A close-up image of a Y-junction shows a small vertex `core' where cracks in any {\it particular} year can interact to approach each other at right angles.  (c) Black and Berg installed pairs of rods here bracketing the crack pattern in 1963.  By 2007 this crack (highlighted by white arrows) has shifted to lie outside the rod pair.  (d) At Victoria Valley, where the terrain is 10-12 ka \cite{Sletten2003}, the  pattern is more hexagonal.}
\end{figure}

More precise results have been gathered through a few long-term field experiments into the dynamics of polygonal terrain.  At Illisarvik, in Arctic Canada, Mackay drained a small permafrost lake, and studied the development of new ice-wedges over the subsequent twenty years \cite{Mackay1984,Mackay2000,Mackay2002}.   The initial crack pattern was strongly influenced by local features, but the most common vertex type was a rectilinear T-junction \cite{Mackay2002}. There were rapid changes in the ice wedges within the first few years, before ceasing due to vegetation growth \cite{Mackay2002}.

In contrast to this, Mackay also established field sites on Garry island, to study well-developed crack patterns \cite{Mackay1974}.  At both locations, he installed cables across the positions of ice-wedges, and measured the time of crack opening as the time at which the cables broke, during several consecutive years \cite{Mackay1974,Mackay1984,Mackay1993,Mackay2000,Mackay2002}.  Not all cracks opened each year, yet cracks could reappear in the same spot after several-year absences \cite{Mackay1993}.  This is similar to drying experiments in clays, where cracks sometimes reappeared after a hiatus of several drying cycles \cite{Goehring2010b}.  The cracks themselves appear to initiate between the position of the buried ice-wedge, and the soil surface \cite{Mackay1984,Mackay2002}.  The interpretation of this is that the wedge itself acts as a memory in this system, as it is preserved over many generations, and is a foreign material that is weaker, or more compliant, than the surrounding soil \cite{Lachenbruch1962,Mackay1984,Mackay2002}.   The cracks advanced slowly, in many cases taking several days to travel from one vertex to the next \cite{Mackay1993}.  It was noted that the cracks could shift positions by several centimetres, in different years \cite{Mackay1974}. The cracks often showed {\it en passant} style doubled paths \cite{Mackay1993}, like the mud cracks shown in Fig. \ref{mudcracks}(d), and other branching structures. It was also seen that cracks initiated at different points, and opened in differing orders, in different years \cite{Mackay1993}.

Another long-term experiment was initiated by Black and Berg in the Dry Valleys of Antarctica \cite{Berg1966,Black1973}, and developed later by Sletten {\it et. al.} \cite{Sletten2003}.  At several sites, pairs of metal rods were driven into the permafrost, bracketing the cracks of polygonal terrain, and the separation of these rods was carefully measured.  An intermittent series of measurements, spanning 40 years, has been made on separation of these rods \cite{Berg1966,Black1973,Sletten2003}, which shows that the intrusive wedges associated with these cracks are growing at rates of $\sim$0.5 mm/year. The observed sites include a range of ages at New Harbor/Taylor Valley (1-2 ka, Figs. \ref{cracks}(d) and \ref{polygons}(a-c)), Victoria Valley (10-12 ka, Fig. \ref{polygons}(d)), and Beacon Valley ($\gtrsim$1 Ma, Fig. \ref{cracks}(e)), which also display different patterns \cite{Sletten2003}.  In older locations the crack pattern is dominated by Y-junctions, while at New Harbour it is closer to a rectilinear patten \cite{Sletten2003}.  Based on the relative ages of these sites, and their degree of ordering, it was suggested that an initially rectilinear pattern of polygonal terrain cracks evolves towards a more hexagonal pattern over thousands of years \cite{Sletten2003}.  

In the vicinity of the New Harbour site, cracks show additional features that agree with the means of evolution proposed here.  As shown in Fig. \ref{polygons}(a), there are regions around many vertices where the cracks curve suddenly to meet the vertex positions, as was sketched in Fig. \ref{model}(c,d), or shown in mud cracks in Fig. \ref{mudcracks}(b,e).  The deviations are consistent with a gradual vertex motion in the direction of the crack that had originally approached at two right angles, similar to that shown for drying mud in Fig. 5(f).  For polygonal terrain, this motion should be recorded in an observable asymmetry of the wedge material beneath the crack pattern;  the deposits on the side that was originally the primary branches of a T-junction should be thicker than the deposits on the side that had approached at two right angles.  This prediction could be checked by excavating a statistically significant number of vertices.

Finally, opportunity was given to revisit the Black and Berg sites in December 2007.  At New Harbour, some vertices showed a complex inner structure, that appears to record how active crack branches curve slightly to intersect each other during different orders of opening in different years, as shown in Fig. \ref{polygons}(b).  Here, it was also noted that cracks at 2 of the 44 pairs of rods installed in 1962 had shifted position to now lie outside the pairs of marker rods (see Fig. \ref{polygons}(c)).  As with Mackay's more direct measurements of crack positions in Arctic polygons, this indicates that the crack positions can shift.  

Taken together, the observations of polygonal terrain show that the fracture network is persistent, but that the pattern does change over time.  The persistence is related to the fact that cracks leave permanent records of their existence within the soil, which preferentially nucleate and guide later generations of cracks.  The order of crack opening also varies, as, for example, not all cracks are active in each year.  All these features were seen mud crack patterns, and were essential to the model of crack pattern evolution developed here.  

\section{Columnar joints}

Columnar joints in lava have attracted scientific curiosity for centuries \cite{RBSRS1693,Huxley1881}.  The same pattern can be observed in other media, such as quenched glass \cite{French1925}, thermally-shocked sandstone \cite{Seshadri1997}, and vitrified ice \cite{Menger2002}.  It also appears in desiccated starch, where it has been particularly well-studied \cite{Huxley1881,French1925,Muller1998,Muller1998b,Muller2001,Toramaru2004,Goehring2005,Mizuguchi2005,Goehring2006,Nishimoto2007,Goehring2009,Crostack2012}.  The geometry of the columns in starch are statistically identical those in lava \cite{Goehring2005,Goehring2008}.  They are, however, usually 0.1-1 cm in diameter, while columns in lava are typically of order 10 - 100 cm across.

In lava, columnar joints form by the gradual extension of thermal contraction cracks into a cooling body, with the cracks growing perpendicular to the isothermal surfaces at the time of cracking \cite{Mallet1875}.  At any particular time, the crack tips represent a network in a thin region near the surface defined by the glass transition temperature isotherm \cite{Ryan1981}, and the columnar forms can be interpreted as a record of this network as it evolved in time, whilst propagating through space.  Lava may cool, and thus columns may grow, in any direction; in many cases both an upper and lower colonnade are present in a single flow unit, as the lava cools from both its top and bottom surfaces (e.g. \cite{Long1986,DeGraff1989}).  In starch, columns grow normal to surfaces of constant moisture content \cite{Goehring2009,Goehring2009b}, and track the position of the so-called funicular-pendular transition, where the water transport mechanisms of capillary flow and vapour diffusion cross over in efficiency \cite{Goehring2009b}.  

As cracks advance in lava, they do not do so continuously, but intermittently, leaving striations on the sides of columns, as shown in Fig. \ref{lava}(e,h), or sketched in Fig \ref{dimless}(a).  Each stria represents a single growth increment: once a crack tip is activated (from some weak point, or vertex), it advances rapidly forward through the brittle lava, and also laterally following near the path of the edge of the crack face's previous increment.  In the forward direction of column growth, however, the tip blunts and stops as it intrudes into a warmer, more plastic zone.  As a result each crack advance leaves a broad thin mark on the side of a column \cite{Ryan1978}. These striae record the cooling conditions of the lava at the time of their formation, and can be interpreted accordingly \cite{Ryan1978,Degraff1987,Aydin1988,Goehring2008}.  In particular, the laterally-averaged height of the striae represents the distance between two isotherms in the cooling lava body, and can be used to measure the cooling rate at the time of fracture \cite{Goehring2008}. 

Striae also record the direction of lateral crack propagation during each small forward advance, in their plumose structure \cite{Degraff1987}.  From this, one can tell whether a crack moved toward, or away from, any vertex in any particular growth increment.  As has been seen elsewhere \cite{Degraff1987,Aydin1988,GoehringPhD}, and as shown in Fig. \ref{lava}(h), this direction can change along subsequent striae.  At one colonnade near Bridge Lake, a site described in Ref. \cite{GoehringPhD}, the directions of growth of 835 striae were measured. Some example growth direction sequences are shown in Fig. \ref{lava}(i). It was found that 64$\pm$2\% of the time, one striae followed the same direction as its predecessor.  While each striae direction is not randomly chosen, the sequence at which cracks are activated at any vertex changes frequently during a column's formation.

\begin{figure}[t]
\includegraphics[width=13.5cm]{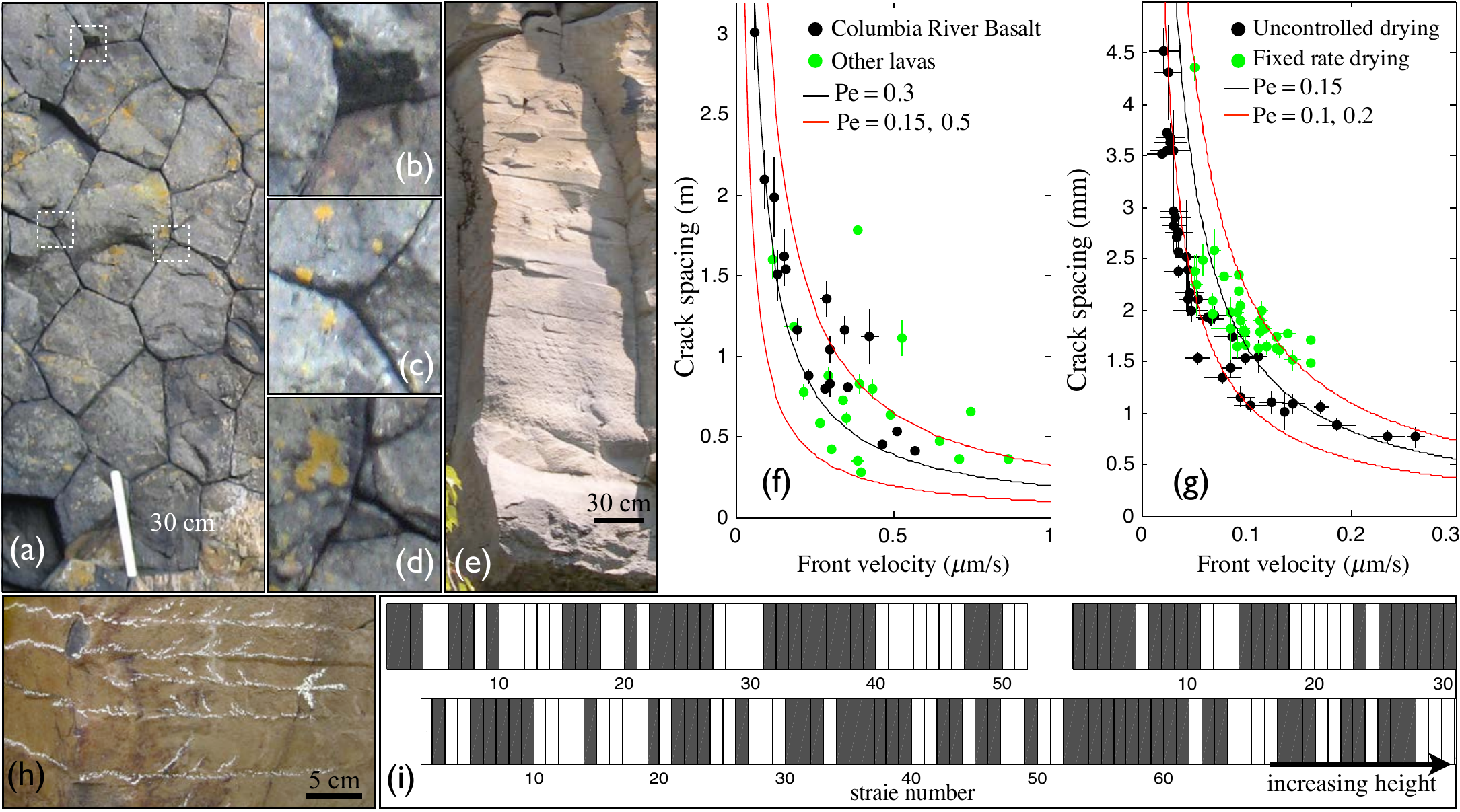}
\caption{\label{lava} Columnar joints in lava. (a) A cross-section of a colonnade near Fingal's Cave, Staffa, Scotland appears hexagonal, but (b-d) cracks very near the Y-junctions curve slightly, reflecting interactions between cracks during any particular growth increment. (e) A single column near Whistler, Canada, showing lateral striae bands.  These can be used to infer the crack front growth velocity \cite{Goehring2008}, which is faster (the striae are smaller) near the base of this flow.  (f) The crack front velocity in basaltic (black) and other lavas (green) is inversely correlated with the size of the columns, as it is also (g) in starch for constant (green) and slowly varying (black) front velocities (panels (f,g) adapted from \cite{Goehring2009}).  Curves show the expected correlation for constant P\'eclet numbers. (h) Plumose, or hairlike marks on striae, here sketched over with chalk, indicate the direction of crack propagation of each crack increment.  (i) The crack growth direction recorded on striae can be toward (white) or away (grey) from any vertex.  Shown here are the direction of crack propagation for sequence of striae on the faces of three columns, near different vertices.}
\end{figure}

The crack pattern at the surface of lava lakes observed in Hawaii are rectilinear in appearance, and dominated by T-junctions \cite{Peck1968,Aydin1988}.  However, the evolution of column vertices was observed at the Boiling Pots, Hawaii, where columns with an average size of $\sim$ 20 cm \cite{Degraff1993} were seen to develop from T-junctions to Y-junctions, over a distance of 1 m \cite{Aydin1988}.  Under careful inspection most vertices were classified there as `pseudo-Y' junctions, as the cracks curved slightly in the immediate vicinity of the vertex \cite{Aydin1988}, similar to the sketch here in Fig. \ref{evolve}(f).  This small deflection was used to argue that cracks did not all interact at a vertex synchronously \cite{Aydin1988}.  In other words, that during each incremental crack advance, there was one crack that arrived at a vertex after the others, causing this slight asymmetry.  Similar vertices can be observed in many colonnades; an example from Staffa is shown in Fig. \ref{lava}(a-d).  In starch colonnades, an evolution from a disordered surface pattern to a hexagonal mature pattern is also observed \cite{Goehring2005}. For columns 1-2 mm across, the pattern orders over a distance of $\sim$1 cm from the drying surface \cite{Goehring2005}; this ordering region can be seen in Fig. \ref{cracks}(b).  Although these observations are sparse, it seems that in both starch and lava the hexagonal columnar pattern is well ordered after 5-10 column diameters of growth.

Thus, columns can evolve by slight variations of the crack plane in each crack increment \cite{Degraff1987}, or by individual cracks overshooting or rounding a vertex formed in the previous cracking cycle \cite{Aydin1988}.  In general, each crack increment is guided by its predecessor, which concentrates stress along its edge, but can advance in a slightly different direction.  In starch this evolution has been followed through x-ray tomography \cite{Muller1998,Goehring2005,Goehring2006,Crostack2012}  and magnetic resonance imaging \cite{Mizuguchi2005}.  In both cases the evolution can be followed for very long times.  Over such times the topology of the network also evolves, and by only three distinct mechanisms: two vertices may collide and exchange position;  crack faces can fail to propagate, merging two (or more) columns; and new columns can initiate from existing vertices \cite{Goehring2006}.  For neighbouring columns of different sizes, there is also a strong tendency to equalise the column area \cite{Goehring2006}.  Taken together, these observations suggest long-term dynamics that are not covered by the model presented here.  

%(or, in other words, the last crack to approach a vertex, during any particular crack growth increment, has curved slightly in response to the earlier cracks)
\begin{figure}[t]
\includegraphics[width=13.5cm]{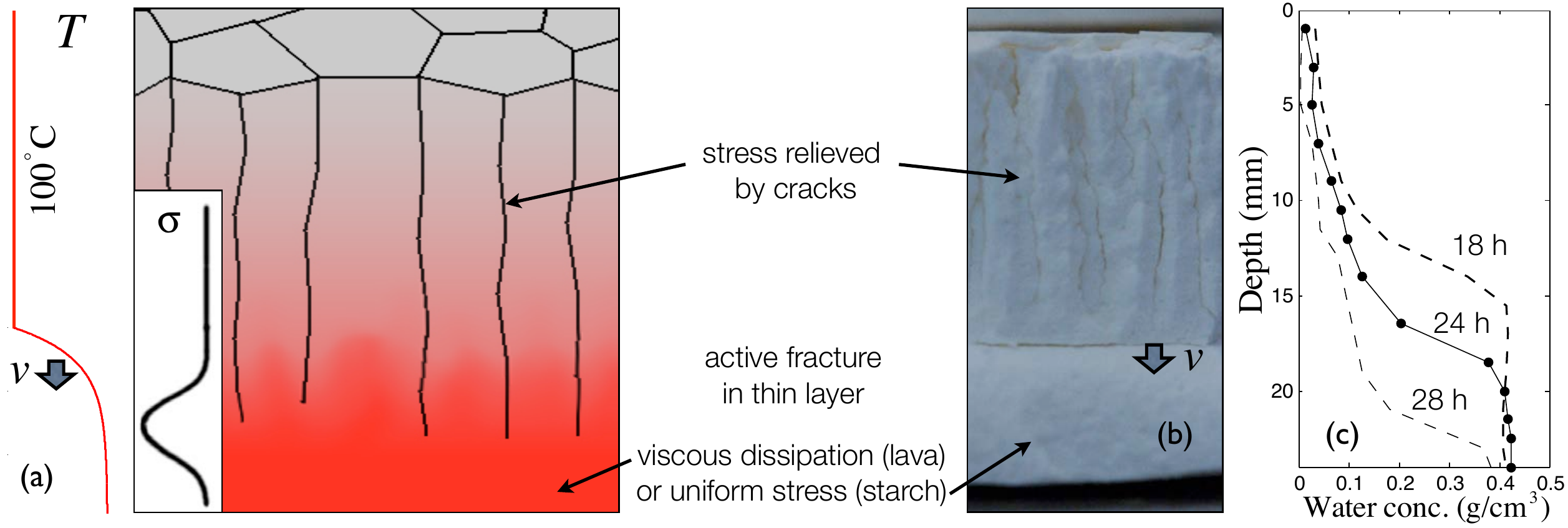}
\caption{\label{dimless} Elastic stresses during columnar joint formation are essentially confined to a thin layer.  (a) In cooling lava, water infiltration within cracks maintains most of a colonnade at 100$^{\circ}$C \cite{Hardee1980}.   A region of steep thermal gradients connects the water-cooled zone with the still-liquid melt, where viscous dissipation relieves stress \cite{Hardee1980}.  The network of crack tips that define the polygons lies within a temperature window of a few degrees, near the glass transition temperature \cite{Goehring2008}.  Each crack face advances intermittently: some point breaks first (yellow dot), causing a crack to advance locally into the warmer lava, where it rapidly halts.  This advance propagates laterally, until it reaches a vertex.  Each such advance leaves behind a single stria \cite{Ryan1978,Degraff1987}, as seen in Fig. \ref{lava}(e,h).  (b) In starch columns grow from the drying surfaces inwards, following (c) a steep transition in water content that generates stress \cite{Goehring2009}. The cooling/drying fronts advance at a rate $v$, that sets a natural length scale $D/v$ to the cracking problem, in place of the film thickness $h$.}
\end{figure}

Next, to consider the scaling of columns, we now need to evaluate the effective thickness of the elastic layer in which they form.  During the solidification of a lava flow, water or moisture in the environment can act an efficient transporter of heat, and the thermal transport problem is approximately that of an arbitrarily thick layer, cooled entirely to 100$^\circ$C, bound to a purely diffusive region over which the temperature transits to that of the internal melt \cite{Hardee1980,Degraff1993,Goehring2008}. Cooling proceeds by the propagation of this solidification front into the lava, with all isotherms traveling together at some velocity $v$, which can be inferred from the average striae heights \cite{Goehring2008}.  In the reference frame of the solidification front, the linear advection-diffusion problem has a steady-state temperature profile $T$, that solves
\begin{equation}
\nabla^2 T + {\rm Pe} \frac{\partial T}{\partial z} = 0
\end{equation}
where Pe = $vL/D$ is the P\'eclet number.  The ratio of diffusion $D$ to advection speed $v$ defines a length scale, which is similar to the film thickness $h$ from Section 2.  As sketched in Fig. \ref{dimless}(a) the thermal gradients are confined to this advective-diffusive layer, as are the elastic stresses. Therefore, according to the arguments presented in Section 2, the crack spacing should be proportional to this effective layer thickness, and inversely proportional to $v$  \cite{Goehring2008}.  More formally, if the length $L$ in the P\'eclet number is identified with the column diameter, then all columns should have the same Pe, which must be of order 1.  As shown in Fig. \ref{lava}(f), this is the case: large columns cooled slower than small ones.

For drying starch, the moisture concentration $\phi$ is functionally equivalent to the temperature, in generating stress.  There is a bottleneck in water transport as the two mechanisms of vapour diffusion and capillary flow cross over in efficiency at some critical moisture concentration \cite{Goehring2009b}.   Most of the stress gradients in a drying starch body are confined to a thin layer, near this bottleneck, where an advection-diffusion problem can be posed for the moisture concentration,
\begin{equation}
\nabla^2 \phi + {\rm Pe_{h}} \frac{\partial \phi}{\partial z} = 0.
\end{equation}
This is demonstrated in Fig. \ref{dimless}(b,c).  Here $D_h$ is the minimum in the moisture diffusivity, and Pe$_h = vL/D_h$ \cite{Goehring2006,Goehring2009}.  As shown in Fig. \ref{lava}(g), the size of columnar joints in starch is inversely proportional to their cracking front velocity $v$, and when scaled by $D_h$ are found at the same P\'eclet number as those in lava \cite{Goehring2009}.

In summary, the spatially extended forms of columnar joints develop in a very similar way to the other hexagonal crack patterns described and modelled here, where the cycles of cracking and healing have become a stacked set of crack advances.  In each advance the network of cracks is guided by the cracks that precede it, but is allowed to take a slightly different path.  The sequence of cracking at any vertex is variable, however, and the vertices themselves interact and move over time.  The crack pattern rapidly evolves from a rectilinear, T-junction dominated pattern to a hexagonal one, but retains a slight asymmetry in the immediate vicinity of vertices that is indicative of the order of crack opening in any particular growth increment.  Finally, as with thin-film crack patterns, the length scale of the polygonal columns scales with the effective layer thickness.

\section{Eroding crack patterns}

Recently, a crack pattern that forms on cemented gypsum sands, on the windward (and therefore eroding) faces of dunes, has been described \cite{Chavdarian2006,Chavdarian2010}, and will be briefly reviewed here.  These crack patterns, shown in Fig. \ref{cracks}(c), are found on the dunes of the White Sands National Monument, in New Mexico. Genetically, they have been argued to be similar to columnar joints, but where the cracks are advancing into the dunes as the cemented sand surface is removed \cite{Chavdarian2010}.    Crack depths of 0.1 cm to 45 cm were reported, and polygon sizes from 5 to greater than 70 cm.  As with naturally-dried desiccation cracks in starch \cite{Muller1998,Goehring2005} and thermal contraction cracks in coke \cite{Jenkins2005}, the polygon size was noted to increase with depth.

It was found that cracks changed directions abruptly as vertices were approached \cite{Chavdarian2010}.  In this, they appear similar to the mud crack experiments reported here and in Ref. \cite{Goehring2010b}, and to the developing polygonal terrain at New Harbour \cite{Sletten2003}.  The angles between cracks at the junctions were measured at 4 sites, and found to deviate significantly from those of orthogonal T-junctions \cite{Chavdarian2010}.  Only in very thin (1 mm) cements, or cases where the cracking directions were strongly influenced by the dune slope, were T-junctions typical.  

A convincing model for the origin of these cracks was also proposed \cite{Chavdarian2010}.  The cycling of moisture in the dunes allows for gypsum to dissolve and re-precipitate near the dune surfaces, cementing the sand.  The same water, as it evaporates, causes stresses which fragment the cemented layer.  Finally, wind erodes the stoss face of the dune, allowing the wetting (and cracking) layer to slowly intrude into the dune.  

Several features of these patterns, such as the local changes in crack orientation near vertices, and the opportunity for dynamics as the face erodes, are consistent with an evolution similar to that outlined here.  However, direct evidence for this is limited.  At one site 13 junctions were excavated, and measured at the surface, 2 cm, and 10 cm depth.  Although it was stated that cracks became straighter, and more ordered, with depth, the distribution of angles from this sample did not significantly change \cite{Chavdarian2010}.  As it is an interesting pattern, potentially related to ordered cracks and water cycling on Mars \cite{Chavdarian2006}, additional work is to be encouraged.  

\section{Summary}

The crack patterns discussed here, such as are commonly seen in dried mud, frozen soils, or old lava flows, can take the appearance of a rectilinear network, with cracks intersecting preferentially at near right-angled T-junctions, or a hexagonal network of equi-angled Y-junctions, or sometimes an intermediate state.  A genetic model was sketched which could account for these observations, amongst others, as the result of simple interactions.  A crack curves to maximise the difference between the strain energy release rate, at the moment of cracking, and the cost of creating the new crack surfaces - an established principle of fracture mechanics \cite{Lawn1993} .  This is a local energy consideration: the crack tip only senses its immediate environment, and responds accordingly.  When a layer repeatedly cracks and heals, the positions of the previous cracks may be weaker, or more compliant, than the surrounding material.  Similarly, in a crack network like that of columnar joints, that is growing through space, the stress state in an un-cracked layer is strongly influenced by the prior pattern.   In this way cracks are guided along the paths taken in previous cracking cycles.  At vertices, however, this argument introduces asymmetries, and a growing crack can be deflected.  If the order in which cracks arrive at a vertex changes with time, in different cracking cycles, these deflections can accumulate and will distort the vertex, moving it in a direction to equalise the angles between all cracks, creating a Y-junction.  The requirements and predictions of this process were outlined in Sections 2 and 3.

This model was compared to and elaborated by presentations of evolving crack patterns in desiccated bentonite clays, polygonal terrain in permafrost, columnar joints in lava and starch, and a pattern of desiccation cracks that evolves as it erodes.   These patterns were found to share similar features, such as a scaling of the crack spacing with the layer thickness, changes in the order of crack opening, and distortions of the crack pattern in the immediate vicinity of vertices.  Furthermore, it was shown in bentonite clay layers that were repeatedly dried and rewetted, that the drift of vertices away from their original positions, during this evolution, scales with the thickness of the cracking layer.  This rate may also depend on the materials involved - as permafrost apparently takes much longer to order than a few tens of cracking cycles, measured there by years.   The direction of this motion was seen to be precisely that predicted to deform vertices towards a hexagonal crack pattern.  

\section{Acknowledgements}

I thank C. Duquette for her assistance in the field studying columnar joints, and R. Sletten, B. Hagedorn, and D. Mann for providing the opportunity to participate in field work in Antarctica, and visit the Black and Berg sites.  I also particularly thank B. Hallet for discussions on this topic over many years, and P. Nandakishore for suggesting an improvement to the algorithms to measure crack angles.

 \end{document}